\documentstyle[12pt,aaspp4]{article}
\lefthead{Crampton et al.}
\righthead{GRAVITATIONAL LENS SBS 1520+530}
 
\begin{document}
 
\title{DETECTION OF THE GALAXY LENSING THE DOUBLY--IMAGED QUASAR SBS 1520+530}

\author{David Crampton\altaffilmark{1}} \affil{Dominion Astrophysical Observatory,
National Research Council of Canada,\\ Victoria, B.C. V8X 4M6, Canada}
\author{Paul L. Schechter} \affil{Department of Physics, Massachusetts Institute of Technology, Cambridge, MA 02138}
\author{J.-L. Beuzit} \affil{Canada-France-Hawaii Telescope Corporation, Waimea, HI}
\altaffiltext{1} {Visiting Astronomer, Canada-France-Hawaii Telescope,
operated by the National Research Council of Canada, the Centre de la
Recherche Scientifique de France, and the University of Hawaii}

\begin{abstract}
$H$ band observations with a spatial resolution of 0\farcs15 carried
out with the Canada-France-Hawaii Telescope adaptive optics system show
a galaxy between the components of the double BAL QSO SBS 1520+530,
thereby confirming this system as a gravitational lens.  The galaxy is
located 0\farcs40 from the fainter of the two QSO images and is offset
0\farcs12 from the line joining them.  The H magnitude of the lensing
galaxy is $\sim$1 mag fainter than expected from the velocity
dispersion derived for the lensing galaxy were it at z = 0.72 or z =
0.81, the redshifts of the two absorption line systems.

\end{abstract}

\keywords{Galaxies: Quasars: individual (SBS 1520+530) -- Gravitational lenses -- Quasars}

\section{INTRODUCTION}

Gravitational lenses have several applications, from the determination
of cosmological parameters, to the measurement of galaxy parameters,
to the detailed study of the inner regions of quasars (see recent
reviews by Narayan and Bartelmann (1998), Refsdal and Surdej (1994)
and Schneider (1996).  The compilation of Keeton and Kochanek (1996)
lists roughly 20 optical quasars and AGN which appear to be multiply
imaged.  While most of the quadruple systems on this list are rated
class ``A'' candidates by Keeton and Kochanek, most of the double
systems are rated class ``B'' or class ``C''.  Two closely spaced
quasars of similar or even identical redshift may or may not be the
result of gravitational lensing; they may simply be a quasar pair
located near to each other in space.  Identification of the lensing
object provides strong additional support that a given candidate is a
true lens.

Recently, Chavushyan et al. (1997, hereafter CVSE) reported that the
QSO SBS 1520+530 has two components, separated by 1\farcs6, with
identical redshifts, z = 1.855.  The two broad emission and absorption line
profiles also appear to be identical, and since BAL type absorptions
are relatively rare, they argue that SBS 1520+530 is very likely to be
a gravitational lens system.

In this paper, we present high spatial resolution observations
demonstrating the presence of a galaxy between the two images,
confirming the CVSE interpretation of the quasar pair
as gravitationally lensed images of a single object.

\section{OBSERVATIONS}

SBS 1520+530 is located 13\arcsec\ from an m$\sim$12 star, which,
though a nuisance for conventional imaging, makes it an ideal target
for imaging with an adaptive optics system. The CFHT Adaptive Optics
Bonnette (AOB) is a general adaptive optics interface which can be
mounted at the cassegrain focus, to which cameras and other
instruments can be mounted (Arsenault et al.  1994, Rigaut et al.
1998). It makes use of a wavefront curvature sensor and a bimorph
mirror with 19 degrees of freedom. For these observations, a
beamsplitter was used that diverted the visible light to the wavefront
sensor while the near-IR light went to the science detector.  The data
were recorded with MONICA, the Universit\'e de Montreal Infrared
Camera, which uses a 256 $\times$ 256 NICMOS HgCdTe array. Special
optics were installed in the camera for use with AOB that yield a
scale of 0\farcs034 per pixel, and a field size of only $9\arcsec
\times 9\arcsec$. Unfortunately, the coatings of these optics had
deteriorated by the time these observations were made and the measured
throughput of the camera was $\sim 4\times$ lower than when initially
installed.

The $H$ filter was selected as offering the best compromise between
sensitivity, good image compensation, and sky brightness. To remove
detector cosmetics and to improve the signal-to-noise ratio, a series
of 180s integrations were made on UT1997 June 14.4 in a four position
dither pattern separated by 3\farcs1 ($\sim$90 pixels), with small
offsets between each set of four exposures. During initial setup, an
additional set of four was also taken at a smaller offset from the
guide star. The region containing lens components A and B received a
total exposure of 48 minutes, but the region near object ``SE" (see
below) only received between 12 and 24 minutes. Measurements of exposures of
the guide star taken immediately after those of SBS 1520+530 indicate
that the AOB was delivering near diffraction-limited images, with FWHM
= 0\farcs13, but measurements of component A on individual 180s data
frames indicate that the seeing was variable throughout the sequence of
exposures.

The raw data frames were median combined in subsets to form sky flats
which were then subtracted from the raw data. Precise offsets were
determined from the positions of point sources in the frames and  these
were also verified using the wavefront sensor coordinates for each
observation. All of the data was then combined with appropriate
(integer pixel) shifts to form the final image. This image, correctly
oriented (see section 3.1), is shown in Figure 1a where the objects are
labelled with the same convention as used by CVSE.  The lensing galaxy
is clearly visible NW of component B.

\section{RESULTS}

Positions and magnitudes of all the objects shown in Figure 1a were
measured on the original, unrotated, image and measurements of the
galaxy were made on images with the point sources subtracted. A variety
of techniques within the {\sc IRAF} package were used in order to
estimate measuring errors. The image of component A was used to
establish the point-spread-function (PSF) which was then scaled and
subtracted from the other point sources with the {\sc DAOPHOT ALLSTAR}
routine. As mentioned above, the FWHM of point sources varied during
the observations and since some parts of the image shown in Figure 1a
were taken at different times, it is not unexpected that the PSFs would
be variable.  In addition, the image quality is expected to degrade
with increasing distance from the guide star due to anisoplanatic
effects in the wavefront correction. Nevertheless, the PSF constructed
from object A appears to be adequate, since the residuals after
subtraction (with {\sc ALLSTAR}) are only slightly larger than those of
the sky background in neighboring areas. Although the NW object appears
to be sharper than objects A,B and SE, the FWHM of point sources on the
final averaged image are all consistent with FWHM = 0\farcs15. Figure 2
shows a portion of the image near the lens before and after subtraction
of a scaled image of component A at the location of B.

\subsection{Astrometry}

Positions of all the objects relative to A are given in Table 1. The
estimated errors in each coordinate are 0.05 pixels or 2mas for the
point sources and $\sim$1 pixel or 30mas for the galaxy. The latter
position was measured after PSF subtraction of the quasar images.
Given the small field, proper orientation of the coordinate system is
not trivial, especially since the MONICA camera is only on temporary
loan to CFHT and hence its mounting orientation is not fixed.
Comparison of positions of stars in M5 obtained during the same run 
with archival HST data indicates that the pixel scale is
0\farcs03416$\pm$0\farcs00003 pixel$^{-1}$ and that north is
$-$100\fdg76 $\pm$0\fdg02 from the vertical columns of the detector.  A
left -- right inversion of the rotated image is also required to
correctly orient the field.  We have also computed the scale directly
from the images of SBS 1520+530, transferring positions from a deep $R$ filter
image obtained at Michigan-Dartmouth-MIT Observatory to the MONICA
image.  We find the same scale and rotation, though with
uncertainties larger by a factor of 3. Portions of the $H$ and $R$ images
are shown on the same scale in Figure 1.

Based on our new calibration, radial coordinates relative to the line
joining A -- B with A as origin are given in columns 4 and 5 of Table
1, and offsets relative to A in equatorial coordinates are listed in
columns 6 and 7. For objects in common, the latter are in good agreement with those listed by CVSE.

\subsection{Photometry}

The relative magnitudes in H of A, B, NW and SE were determined with DAOPHOT
using a PSF (determined from image A) which extended to a radius of 25
pixels or 0\farcs85. These are listed in the last column of Table 3.
The average error reported by DAOPHOT was 0.05mag for the point
sources. Since these magnitudes were measured in an internally
consistent way, their accuracy relative to each other is quite good.
However, observations of only one standard star were taken during the
run, three nights later, due to poor weather after these lens
observations were obtained.  The observation of FS30 (Casali and
Hawarden 1992) yields a zeropoint of $H$ = 17.4$\pm$0.1 mag for
component A. Lensing is achromatic, so the magnitude differences
between objects should be identical, within errors, (or variability on
scales short compared to the time delay), to those observed in $V$ by
CVSE if they are lensed components. This is true for objects B and NW,
but SE is much redder.

While our $R$ data have not been photometrically calibrated, we find
that the B+G component is 0.60 mag fainter than the A component
and that the NW object is 1.25 mag fainter.  This is roughly
consistent with the CVSE $R$ photometry.  Fitting separately for
the brightness of the lensing galaxy by fixing its position relative
to the quasar images (see Schechter and Moore 1993) we find that B is
0.66 mag fainter than A and that the galaxy is $3.7 \pm 0.4$ mag
fainter.  Based on the CVSE photometry for A, the galaxy then 
has $R$ = 21.6. 

The relative H magnitude of the galaxy was measured by aperture
photometry to the same radial distance, 25 pixels, as the point
sources. Using the above calibration, the galaxy has $H$ = 19.2. CVSE
noted the presence of two absorption line systems in the spectra of SBS
1520+530, at z = 0.716 and z = 0.815. Assuming that the lensing galaxy
is responsible for one or other of these, we adopt z = 0.765 as a
plausible lens redshift for the purpose of interpreting our
observations.  Adopting H$_0$ = 50 km s$^{-1}$ Mpc$^{-1}$ and $\Omega =
1$ we find a restframe absolute magnitude M$_{AB}(9350)$ = $-$22.4 on
the Oke system.  For a typical gE galaxy, this is equivalent to M$_B$ =
$-$20.4 or nearly a magnitude fainter than an L$^*$ galaxy. The
observed $(R-H)$ color ($\sim2.4\pm0.4$) is much bluer than expected
for an elliptical galaxy, being appropriate for an Scd galaxy at the
assumed redshift. We stress, however, that the galaxy color is quite
uncertain without higher spatial resolution observations in the visible
wavelength region and, furthermore, that its redshift has not yet been
measured.

\section{MODELS}

The non-collinearity of the lensing galaxy and the two images is a
clear signature of the non-axisymmetric character of the lensing
potential.  A minimum of 5 parameters is needed to describe such a
potential: a strength, which governs the splitting, the two source
coordinates, which are not observable quantities, and an amplitude and
orientation for a shear which breaks the symmetry.  A simple and
plausible model incorporating these features is the singular isothermal
quadrupole potential (e.g Kochanek 1991)
$$ \psi = br[1 +\gamma \cos2(\theta - \theta_\gamma)] $$ 
where $\psi$ is the projected potential
(cf. Narayan and Bartelmann 1996), $b$ is the lens strength (an angle),
$r$ and $\theta$ are polar coordinates on the sky measured from the
center of the potential (where $r$ is an angle), $\gamma$ is the
dimensionless shear and $\theta_\gamma$ gives the orientation of the
shear.  The linear dependence of the leading term on $r$ would give a
galaxy with a constant circular velocity.  The linear quadrupole term
gives isopotentials which are self-similar and approximately
elliptical, as  might be expected from an elliptical isothermal mass
distribution with an ellipticity $\sim 6 \gamma$ in the limit of small
shear.

The positions of the two quasar images relative to the lensing galaxy
(which we take as the origin of our coordinate system) give us two
constraints each, leaving us with one constraint less than needed for
a 5 parameter model.  The obvious choice for a fifth constraint is the
observed flux ratio of the two quasar images.  There are several cases
for which observed magnifications fail to agree with otherwise
excellent models, e.g. MG0414+0534 (Witt et al. 1995) and B1422+231
(Mao and Schneider 1997), for which microlensing and millilensing have
been invoked to explain the discrepancies.  But as Mao and Schneider
note, magnifications are more likely to be reliable when they are
relatively small, as in double systems, than when they are larger, as
in quadruples.

Our model yields a strength $b = 0\farcs77$ and a shear $\gamma = 0.066$
with $\theta_\gamma = 91\fdg2$ measured E from N.  This corresponds to
an E3-4 galaxy with major axis at PA $1\fdg2$.  The value of $\gamma$
is determined largely by the magnification constraint.  The orientation
and ellipticity of the observed galaxy is difficult to measure with
precision given its relative faintness and the presence of component
B.  On the B-subtracted image (Fig. 2), the brighter contours appear to
be oriented almost N--S, but the fainter contours give a PA =
$-20\pm17^\circ$.  The derived amplitude and direction of the shear are
thus consistent with our observations, but a better image is required
in order to determine whether it is solely responsible. In a recent
survey of the parameters of gravitational lens galaxies, Keeton,
Kochanek \& Falco (1997) found excellent agreement between the observed
and model position angles in 14 out of 17 cases. It appears likely that
the galaxy lensing SBS 1520+530 is another such case.

We also investigated the possibility that the shear might arise from a
tide due to a neighboring galaxy or galaxies.  Adopting Kochanek's
(1991) external shear model we find the position angle unchanged, but
with $b = 0\farcs73$ and $\gamma = 0.13$.  From our deep MDM data, the
nearest object which might be another galaxy at the same redshift as
the lens is an $R \sim 22.6$ object 1\farcs88 seconds from the lensing
galaxy at P.A.  50$^\circ$ (see Fig. 1b). Since its position angle is
very different from that derived it is unlikely to be a significant
component of the computed shear.

The one dimensional velocity dispersion $\sigma$ for our isothermal
model is given by $$ {\sigma^2 \over c^2} = {D_S \over D_{LS}}{b \over
4 \pi} $$ where the angular diameter distances are to the source and
the lens.  Taking $z_L = 0.765$ we find a velocity dispersion of 255 km
s$^{-1}$, appropriate to an elliptical galaxy 0.5 mag brighter than L*
(Fukugita \& Turner 1991). However, as noted above, the measured
brightness of the lensing galaxy is $\sim$1 mag fainter than this,
implying the presence of some additional convergence. One is tempted to
conclude either that the lensing galaxy is underluminous by a factor of
four, or that there is additional mass present (e.g., diffuse cluster
matter projected onto the lensing galaxy).  Keeton et al. (1997) study
a large sample of lenses and also find mass-to-light ratios which are
somewhat larger than expected from stellar dynamics, though with
considerable scatter.

For an isothermal sphere the expected time delay in the absence of
shear is given by 
$$ \tau_{AB} = {1+z_L \over c } {D_L D_S \over
D_{LS}} {1 \over 2} ({r_B}^2 - {r_A}^2)$$ 
where $D_{LS}$ is the angular
diameter distance from the lens to the source and where the positions
of the images A and B are in radians measured from the lensing galaxy.
This gives an expected time delay of $1.87 \times 10^{-11}  H_0^{-1}$, of
the order of several months.  Corrections of order $\gamma$ are needed
when the shear is taken into account.

\section{SUMMARY}

High spatial resolution near-IR observations reveal the presence of a
galaxy between the quasar pair SBS 1520+530, thereby confirming it as a
gravitational lens system. The galaxy is substantially offset from the
line joining the two lensed images, implying the presence of shear
which our data indicate may arise from the orientation of the lensing
galaxy. The galaxy appears to be fainter than predicted
by a straightforward lens model.

The presence of the bright star close to SBS 1520+530 means that it is
an excellent candidate for future monitoring with adaptive optic
systems. High spatial resolution images and spectra of the lensing 
galaxy are required to provide additional constraints on the lens
model.

\acknowledgments
   We thank Pierre Couturier and Dennis Crabtree for making CFHT Director's
time available for this project, and John Hutchings, Eric Steinbring, Jim
Thomas and all the members of the AOB team for their assistance with
the AOB instrument and observations.  PLS gratefully acknowledges
support from NSF grant AST96-16866.

\clearpage

\begin{table}
\caption[]{Positions and Magnitudes Relative to Component A}
\begin{tabular}{crrrrrrr}

 Object & x~~~ & y~~~ & r~~~ & theta$^1$ & RA~ & DEC & $\Delta H$ \\ &
pix~~ & pix~~ & $\arcsec$~~~ & $\deg$~~ & $\arcsec$~~~ & $\arcsec$~~~ & mag \\
\hline
  B  & +10.95 & +44.61 &  1.569 &   0 &    1.427 & $-$0.652 & 0.69 \\
GAL  &  +4.85 & +34.62 &  1.194 &  5.82 &  1.131 & $-$0.384 & 1.75 \\
 NW  & $-$11.24 & $-$77.23 & $-$2.666 &  5.51 & $-$2.520 & 0.870 & 1.31 \\
 SE  & +58.79 & +112.52 & 4.337 & $-$13.80 &  3.401 & $-$2.691 & 0.38 \\

\end{tabular}
\tablenotetext{1}{Angle measured counterclockwise from A -- B direction}

\end{table}
\clearpage

\begin{figure}
\plotone{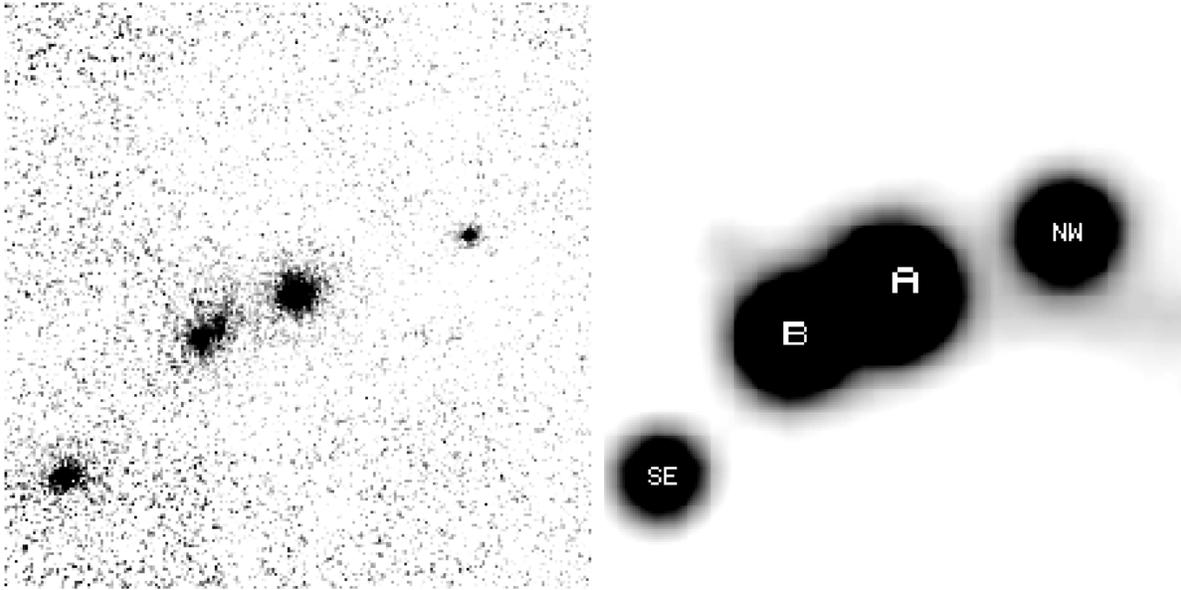}
\figcaption[crampton.fig1.ps]
{(Left; Fig. 1a) Median-combined images of the SBS 1520+530
field in the $H$ band. Due to differing numbers of exposures in
different parts of the image, the noise characteristics vary. The
lensing galaxy is visible NW of component B.  (Right; Fig. 1b) The
$R$ image of the same field.  The seeing was FWHM = 0\farcs8 but the
display was scaled to show the faint object to the NE of component B.
In both figures N is up and E is to the left and the panels are 8\farcs5 on a side.  }
\end{figure}

\begin{figure}
\plotone{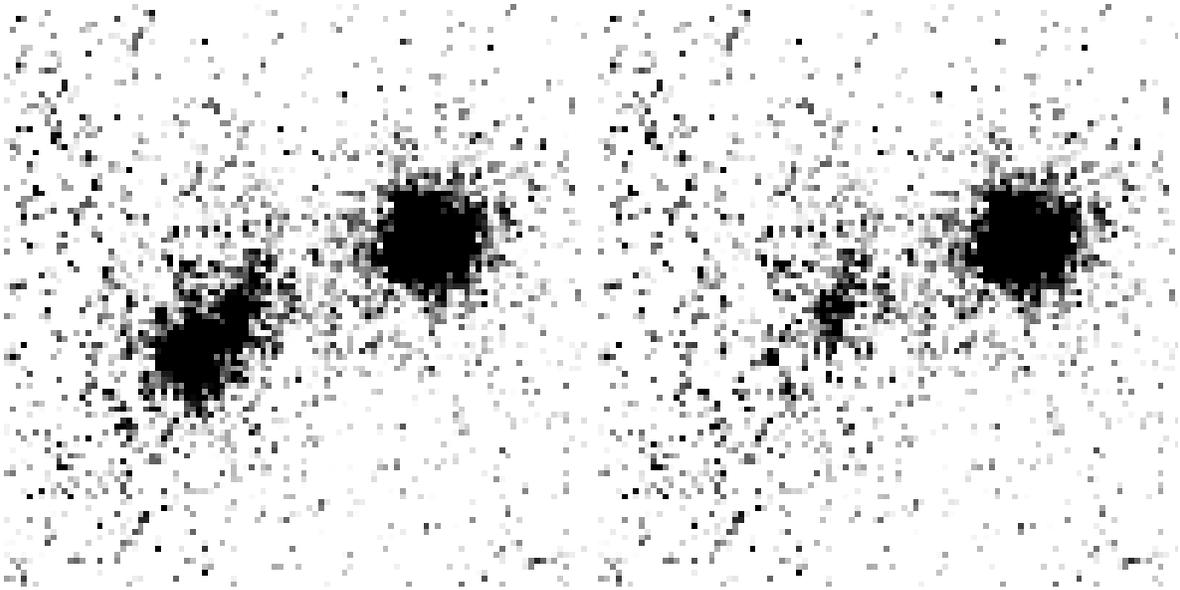}
\figcaption[crampton.fig2.ps]
{(Left) Enlargement of the $H$ image near components A and B. (Right)
Same but with component B removed by subtracting a scaled image of A to 
better show the lensing galaxy. The panels are 3\farcs4 on a side.}
\end{figure}

\end{document}